\documentclass[published]{JHEP3} 

\JHEP{00(2007)000}

\JHEPspecialurl{http://jhep.sissa.it/JOURNAL/JHEP3.tar.gz}

\usepackage{epsfig,multicol,bbm}

\usepackage{graphicx}
\usepackage{dcolumn}
\usepackage{bm}

\newcommand\fverb{\setbox\fverbbox=\hbox\bgroup\verb}
\newcommand\fverbdo{\egroup\medskip\noindent%
			\fbox{\unhbox\fverbbox}\ }
\newcommand\fverbit{\egroup\item[\fbox{\unhbox\fverbbox}]}
\newbox\fverbbox

\newcommand{\be}{\begin{equation}}
\newcommand{\ee}{\end{equation}}
\newcommand{\ba}{\begin{eqnarray}}
\newcommand{\ea}{\end{eqnarray}}

\newcommand{\err}{\end{array}}
\newcommand{\bc}{\begin{center}}
\newcommand{\ec}{\end{center}}

\newcommand{\hmpc}{h$^{-1}${\rm Mpc}}

\newcommand{\etal}{et al.}
\newcommand{\eg}{e.g.,}
\newcommand{\ie}{i.e.,}
\newcommand{\rms}{r.m.s.}
\newcommand{\lcdm}{$\Lambda$CDM }

\newcommand{\prtl} {\partial}
\newcommand{\veck} {{\bf k}}
\newcommand{\vecn} {{\bf n}}
\newcommand{\vecq} {{\bf q}}
\newcommand{\vecr} {{\bf r}}
\newcommand{\vecs} {{\bf s}}
\newcommand{\vecx} {{\bf x}}

\newcommand{\del} {\delta}

\newcommand{\aphj}{{\it Astrophys. J.}}
\newcommand{\mn}{{\it Mon. Not. R. Astron. Soc.}}
\newcommand{\asta}{{\it Astron. Astrophys.}}

\title{The origin of  'Great Walls'}

\author{Sergei F. Shandarin\\
            Department of Physics and Astronomy, 
            University of Kansas, KS 66045, U.S.A.\\
            E-mail: \email{sergei@ku.edu}}

\received{\today} 		
\accepted{\today}		

\preprint{\hepth{???}}	

\abstract{A new semi-analytical model that explains  the formation and sizes of the 'great walls' - 
the largest structures observed in the universe is suggested. Although the basis of the model is the 
Zel'dovich approximation it has been used in a new way very different from the previous studies. 
Instead of traditional approach that evaluates the nonlinear density field it has been utilized for  
identification of the regions in Lagrangian space that after the mapping to real or redshift space 
(depending on the kind of structure is studied) end up in the regions where shell-crossing occurs. 
The set of these regions in Lagrangian space form the progenitor of the structure and after
the mapping it determines the pattern of the structure in real or redshift space. The particle trajectories have crossed in such regions and the mapping is no longer unique there. 
The progenitor after mapping makes only one stream in the multi-stream flow regions therefore 
it does not comprise all the mass. Nevertheless, it approximately retains the shape of the structure. 
The progenitor of the structure in real space is determined by the linear density field along
 with two non-Gaussian fields derived from the initial potential.  Its shape in Eulerian space is also
affected by the displacement field. The progenitor of the structure in redshift space also depends 
on these fields but in addition it is strongly affected by  two anisotropic fields that determine the pattern of great walls as well as their huge sizes.  All the fields used in the mappings
are derived from the linear potential smoothed at the current scale of nonlinearity which is
$R_{nl} = 2.7$ {\hmpc} for the adopted parameters of the \lcdm universe normalized to $\sigma_8 = 0.8$.  The model predicts the existence of walls with sizes significantly greater than 500 {\hmpc}
that may be found in sufficiently large redshift surveys. }


\begin{document}

\section{Introduction}
\label{sec:intro}
The first indications of the existence of superclusters of galaxies bridging the 
clusters of galaxies \cite{gre-tho78,chi-roo79} was confirmed by the
CfA redshift survey  \cite{del-etal86}. In the following two decades 
the discovery of huge concentrations of galaxies spanning over 200 {\hmpc}  were 
reported   ({\eg} \cite{gel-huc89,got-etal05}).
They were dubbed  "great walls" due to their sizes and geometry in redshift space.
The current record belongs to
"a Sloan Great Wall of galaxies 1.37 billion light years long, 80\% longer than 
the Great Wall discovered by Geller and Huchra and therefore the largest observed 
structure in the universe" \cite{got-etal05}.  Taking into account that the SDSS
redshift survey is much larger and deeper  than CfA one may wonder whether
the Sloan Great Wall will remain the greatest wall or even greater walls will be found in 
the future redshift surveys and if it is so how large it may be.
The answer to this question may depend to certain extent on the exact definition of the walls. 
For instance, some definitions may result in the percolating system of filaments and walls
that would span throughout the whole volume.  
However even in this case the essential constituents  can be probably identified and  measured.
We show that by restricting the analysis to the issues of overall geometry and scale 
of the largest walls one can alleviate some of the problems of this kind and make some progress.
In particular, the details of the density distribution within the walls, such as the
accurate positions and masses of halos become less important if the scales of 
interest exceeds $\sim$10 {\hmpc}.

We begin with a brief  discussion of the structure in real space. The scale
separating the nonlinear regime of the gravitational growth 
from  linear or mildly nonlinear regimes  is roughly  around 5  {\hmpc} which is close
to the galaxy correlation scale. It seems  to be by far too small to be directly 
relevant to the scale of filaments some of them are a least 70 -- 100 {\hmpc} long. 
Bond, Kofman and Pogosyan \cite{bon-etal96}  (hereafter BKP) suggested that the filamentary network 
in real space with a typical scale of $\sim$30 {\hmpc}
can be explained  by invoking the correlation bridges between relatively 
high and therefore rare peaks in the linear density field filtered on scales greater than
$R_{\rm b}$ such that $\sigma_{\rho}(R_{\rm b}) \lesssim 1$. 
Unfortunately,  the BKP model has not provided a framework for a 
quantitative evaluation  of the density contrast in the filaments bridging the clusters
of galaxies from the enhancement of the conditional correlation function between 
two or more peaks. 
Besides, the gravitational growth of the structure in the universe is a deterministic 
process if the linear perturbation field is specified. 
Therefore, one may seek a deterministic  relation between the initial field and 
 the final structure at least in  cosmological simulations where the full information 
 is available. 
In addition, the cluster analysis of the nonlinear dark matter density field in real space
obtained by N-body simulation in the \lcdm model   \cite{virgo98}
revealed the filaments with lengths up to $\sim$100 {\hmpc} \cite{sh-she-sah04}
{\ie} significantly greater than 30 {\hmpc}.
A similar analysis of the mock galaxy catalogs  \cite{col-etal98} demonstrated the presence of even 
longer filaments in redshift space, up to $\sim$ 150 {\hmpc}  \cite{she04}. Both numbers 
are almost certainly  affected by the simulation box sizes (240 {\hmpc} in the former and 346 {\hmpc} 
in the latter case) and could be even longer in larger boxes. 
The simulated dark matter density fields and especially mock galaxy catalogs
look similar to the observed distribution of galaxies as well as 
they have similar statistics of various kinds that makes the model viable.
However, the emergence of the nonlinear structures (great walls) $\sim$ 50 -- 100 times 
greater than the scale of nonlinearity remans unexplained even in the cosmological N-body simulations 
where the full information about the both initial conditions and formed structure is available. 

It has been realized for a long time that the structures in real space and 
observed in redshift space have different pdfs, power spectra, correlation functions,
and higher order moments 
({\eg} \cite{pee80, kai87,hiv-etal95,sza-etal98,hui-kof-sh00,pea-etal01}). 
The redshift-space correlation function has an oblate shape on scales greater than a few {\hmpc} 
indicating the presence of anisotropic structures in redshift space flattened in the radial 
direction. However, the
measurements do not go beyond $\sim$ 30 {\hmpc} where the correlation function drops to
$\sim 0.1$ and the signal drowns in  noise \cite{pea-etal01}.
Theory has not made any specific predictions concerning  the structure in the universe on scales 
greater than 100 {\hmpc} therefore the conventional wisdom is that  the perturbations on such 
scales must be in the linear regime. 

The 'finger of God' is the most conspicuous and best understood effect in redshift space.
However, it is a relatively small scale effect and cannot explain the coherence of the great walls 
over a few hundred {\hmpc}. 
It has been also demonstrated in \cite{pra-etal97,mel-etal98} that the size of the structures in the 
redshift space is considerably greater than that of the parent structures in real
space.  In addition, the authors emphasized a characteristic circular
pattern in redshift space and the increased 
spacing between the structures in the redshift direction.  However, they neither  mentioned the 
enlargement of sizes of the structures in the transverse direction nor provided explanation 
to their observations.
  
The presence of walls up to 150 {\hmpc} in length 
in redshift space in three-dimensional simulations based on the adhesion
approximation was emphasized  in \cite{wei-gun90}. 
The authors pointed out that the correlation function was weak or even negative at these scales. 
They also observed that
$\sim$ 100 {\hmpc} walls grow from several favorably aligned high-density peaks 
in the linear density field, each coherent over $\sim$ 20 {\hmpc}. This seems to be a factual
description of the process, however similarly to the BKP model it requires a dynamical model 
different from the linear theory of gravitational instability.
The major problem with the structures spanning over $\sim$ 300 {\hmpc} arises
from a lack of a natural scale of this magnitude in the linear density field. If the
field is smoothed with say 100 {\hmpc} or greater scale than its amplitude 
becomes so low that such a field would remain almost perfectly linear 
at the present time. 

We view the formation of the structure as a continuous mapping from 
the initial practically uniform state in Lagrangian space to the final highly inhomogeneous and anisotropic 
distribution of mass on scales up to $\sim$ 100 {\hmpc} in Eulerian space followed by another mapping
to redshift space where the largest structures become greater than 300 {\hmpc}. 
These mappings involve the transport of mass (although not physical in the case of the mapping to
redshift space), however the rms displacement of mass elements is only about 15 {\hmpc}
and therefore it cannot produce the structures of needed sizes by itself.  
Thus, the certain initial fields {\ie} the fields present in the linear stage and playing an important role 
in the nonlinear dynamics must have scales in the range of 100 {\hmpc} or even greater.
The major goal of this paper is to show that the fields having needed properties exist 
in Lagrangian space.
The outcome of their combined effect is the definition 
of two parent structures that are the progenitors of the large-scale structures: 
one in real and the other in redshift space. Being closely related two progenitors are still quite distinct. 
The properties of the progenitors are determined by a  combination of Gaussian and 
non-Gaussian fields derived from the initial gravitational potential smoothed with a Gaussian filter 
at the scale of nonlinearity 
$R_{\rm nl}$ defined by the equation  $<\delta^2_{\rm lin}(R_{\rm nl})> =1$, where 
$\delta \equiv \delta \rho/\bar{\rho}$.
In  the standard \lcdm model normalized to $\sigma_8 \approx 0.8$  
this scale is $R_{\rm nl}$ = 2.7 {\hmpc}. The choice of the filter and filtering scale 
for 2D illustrations approximately corresponds to an optimal choice of the filter and scale
in 3D simulations that studied the accuracy of the Truncated Zel'dovich Approximation
(TZA) in a set of power law models in the Einstinen-de Sitter model \cite{mel-pel-sh94}. 
Some adjustment will be probably needed for the \lcdm model in 3D. Based on that study
I would conservatively guess that the change in $R_{\rm nl}$ would not be greater than 30\%.
No fields smoothed at greater scales are involved in the mapping.
The structure resulting from the mapping of the progenitors cannot and is not supposed 
to reproduce accurately the mass or galaxy number densities, 
however the overall shape and sizes of the structure in real space and great walls in
redshift space must be depicted quite well. 

We use TZA as a mapping device
for both mapping \cite{col-mel-sh93,mel-pel-sh94,mel-sh-wei94,hui-kof-sh00}. 
In this paper we will
concentrate on the lengths  of the progenitors in the Lagrangian
space and their transformation caused by the mapping to Eulerian and redshift space. 

We illustrate the model by two-dimensional plots. Although no two-dimensional
model can reproduce all the properties of a three-dimensional system it may provide
a reasonably good description of major ideas.
In order to make the appearance of illustrations more 
realistic we use the linear power spectrum $P^{2D}(k) = k\,P^{3D}(k)$, where
$P^{3D}(k)$ is the power spectrum of the linear density perturbations in the \lcdm model
with parameters  $h=0.7$, $\Omega_m=0.3$, $\Omega_b = 0.047$, $n=1$.
We use the BBKS transfer function \cite{bar-etal86} with the $\Gamma$-parameter
given by \cite{sug95}. This choice of  two-dimensional power spectrum retains
all the moments $<k^j>$ of the three-dimensional spectrum and therefore 
all the scales of the two-dimensional field determined by the ratio of two moments
({\eg}  the scale of peaks $R_*^2 \sim <k^2>/<k^4>$) remain similar to that of  the  
three-dimensional field except a small factor $\sqrt{3/2}=1.2$.
The two-dimensional fields were generated in 1024 {\hmpc} box on 1024$^2$ mesh.
 
 The rest of the paper is organized as follows. Section 2 outlines the dynamical model
 and fields involved in both mappings. Section 3 provides two dimensional illustrations that
 elucidate the main concepts of the model, the axes in the figures showing various fields are
 given in {\hmpc}. Section 4 discuss the differences between 2D
 and 3D cases, Sec. 5 summarizes the results and outline the further problems, finally
 the appendix gives the joint density probability function of three invariants of the deformation
 tensor that are used in the mappings. 
\section{Model}
The dominant feature of the process of the structure formation is usually dubbed
as the hierarchical clustering. It simply means that small halos generally form 
earlier than the larger ones.  
Thus, more  massive halos are formed by consecutive merging of smaller
halos. The merging process effectively stops on the scale of $R_{\rm nl}$  or on a 
little greater scale.  It is a well established fact that he formation of the structures on 
larger scales is mainly determined by the long wave part of the initial spectrum 
with $k < R^{-1}_{\rm nl}$
\cite{lit-wei-par91}.  \cite{lit-wei-par91} also pointed out that the changes
in the high-frequency ($k > R^{-1}_{\rm nl}$) components of the initial conditions 
did not affect much the evolution of long waves. A direct comparison of the nonlinear 
density fields obtained in the N-body simulations of models with and without high-frequency
components (Truncated Zel'dovich Approximation) in the initial conditions has also shown 
a very good agreement \cite{col-mel-sh93,mel-pel-sh94,mel-sh-wei94}. In general, the
Zel'dovich approximation
"gives a reasonable description of the overall pattern of N-body simulations" \cite{bon-etal96}. 

Using the comoving coordinates $\vec x \equiv \vecr / a(t)$ where $a(t)$ is the scale factor
describing the uniform expansion of the universe the Zel'dovich
approximation \cite{z70} (see also \cite{sh-z89})  becomes
\be
x_i(\vecq, t) = q_i + D(t)s_i(\vecq),
\label{eq:zappr}
\ee
where  vector  $\vecq$ gives the Lagrangian (initial) position of a fluid element,
$\vecx$ is the Eulireian position of the fluid element at time $t$,  
$D(t)$ describes the linear growing mode of fluctuations, 
$s_i(\vecq) = - \prtl \Phi/\prtl q_i$ is the displacement field and  $\Phi(\vecq)$ 
is the linear gravitational potential field. Thus, our
dynamical model is the Zel'dovich approximation applied to the initial fluctuation field smoothed
over the nonlinear scale $R_{\rm nl}$ which equals 2.7 {\hmpc} in the {\lcdm} model normalized
to $\sigma_8 = 0.8$. 

Instead of traditional approach that tries to evaluate the evolution of density field 
we suggest to consider its inverse {\ie} the evolution of specific  volume $V = \rho^{-1} $. 
The specific volume is more 
suitable quantity for our purpose because it allows an easy separation of the effects 
related to various fields.
We would like to remind that we do not use this model
for accurate predicting of the nonlinear density field but only for approximate 
identification of high density regions. 
The specific volume is given by  the Jacobian
\be
{V_{x}(\vecq,t) \over <{V}>} = J_{x} = \det\left({\partial x_i \over \partial q_j}\right).
\label{eq:vx}
\ee
The growth of density corresponds to the decrease of the specific volume.
Some fluid elements may collapse to zero volume and  at later time 
their volumes predicted by eq. \ref{eq:vx} become even negative
however this is an easy  problem. 
Negativeness of the density or specific volume  simply reflects the fact that such 
a fluid element turns inside out and  the positiveness of the density can be easily restored by 
taking the absolute value of the Jacobian.  It also indicates that the trajectories of fluid elements 
are crossed and the mapping in no longer unique. This creates a much more serious problem
that does not have a simple solution even in the frame of the Zel'dovich approximation
(apart from breaking the approximation within the regions where shell-crossing occurs) 
mainly because the problem becomes nonlocal. 
The set of fluid elements that collapse by the present time form a  structure in Lagrangian space,
we shall call it the progenitor or parent structure. 
The parent structure mapped to Eulerian space by means of eq. \ref{eq:zappr} forms 
the pattern of the large-scale structure.  It is worth stressing that the parent elements do not represent the total mass contained in the large-scale structure however they approximately 
mark the regions of highest density.  The situation becomes more complex if we recall that
the large-scale structure forms from highly nonuniform state where practically all the
mass is already in the form of virialized halos. However, as the previous studies showed
( {\eg}\cite{sh-z89,lit-wei-par91,col-mel-sh93,mel-pel-sh94,mel-sh-wei94}) the overall
distribution of the halo number density  corresponds to this model quite well. 

The Jacobian can be expressed in terms of the invariants ($I_1, I_2, I_3$) or eigen values
($\lambda_1, \lambda_2, \lambda_3$) of the deformation tensor 
 $d_{ij} = - \prtl s_i/\prtl q_j = \prtl^2 \Phi/\prtl q_i \prtl q_j$
\ba
J_{x}(\vecq, t) &=& 1-D(t)I_1(\vecq)+D^2(t) I_2(\vecq) - D^3(t) I_3(\vecq) \nonumber \\
&=& \left[1-D(t) \lambda_1(\vecq)][1-D(t)\lambda_2(\vecq)][1-D(t)\lambda_3(\vecq)\right],
\label{eq:Jx3}
\ea
where 
\be
I_1=d_{11}+d_{22}+d_{33}, ~~
I_2= M_{11}+M_{22}+M_{33} ,~~
I_3= \left| \begin{array}{c c c}
 d_{11} &  d_{12} & d_{13}\\
 d_{12}  & d_{22} & d_{23}\\
 d_{13}  & d_{23} & d_{33}
 \end{array} \right|,
 \label{eq:invariants}
\ee
and $M_{ij}$ are the minors of the corresponding elements of the determinant $I_3$.
The eigen values are the solutions of the characteristic equation
\be
\lambda^3 - I_1 \lambda^2 + I_2 \lambda - I_3 = 0.
\label{eq:charact}
\ee
It always has three real solutions because the mapping is generated
by the potential vector field $s_i({\bf q})$ and therefore the deformation tensor is symmetric
$d_{ij}=d_{ji}=\partial^2 \Phi / \partial q_i \partial q_j$; 
it is convenient to assume  the solutions to be ordered at every point 
$\lambda_1(\vecq) \ge \lambda_2(\vecq)$ and $\lambda_2(\vecq) \ge \lambda_3(\vecq)$.

Equation $J_x(\vecq,t) =0$ has obviously three solutions for $D$ as well:   $D_i = 1/\lambda_i$.
If any of them lies in the interval $0 < D_i(\vecq) <1$ we consider the fluid element with Lagrangian
coordinates $\vecq$ to belong the progenitor of the structure. Condition $0 < \lambda_1(\vecq) < 1$
is equivalent to the above condition for $D_i(\vecq)$ but it is less general and cannot be used
for the mapping to redshift space where it is not of potential type.
 
In order to describe the mapping to redshift space ($\zeta_i$) we  supplement eq. \ref{eq:zappr}
with additional term 
(see {\eg} \cite{hui-kof-sh00})
\begin{equation}
\zeta_i = x_i + f D s_k(\vecq) n_k(\vecx) n_i(\vecx) = q_i + Ds_i(\vecq) + f D s_k(\vecq) n_k(\vecx) n_i(\vecx)
\label{eq:qtoz3}
\end{equation}
where  $\vecx$  is given by eq. \ref{eq:zappr}, 
$f = d\log D/ d\log a \approx \Omega_m^{0.6}$ for the present time \cite{pee80} 
or a better approximation  given in \cite{car-etal92}.
The unit vector $n_i \equiv x_i/x$ is  directed along the line of sight. 
We assume summation over repeated indices throughout the paper.  
The peculiar velocity $v^{(p)}_i =\dot{a}fDs_i$ and therefore the displacement of a galaxy
from comoving Eulerian position, $x_i$ to redshift space position, $\zeta_i$,
 is $v^{(p)}_k n_k /(aH)= f D s_k n_k$ along the line of sight. The distances in Lagrangian,
 comoving Eulerian  and redshift space are measured in {\hmpc}.
\FIGURE{\epsfig{file=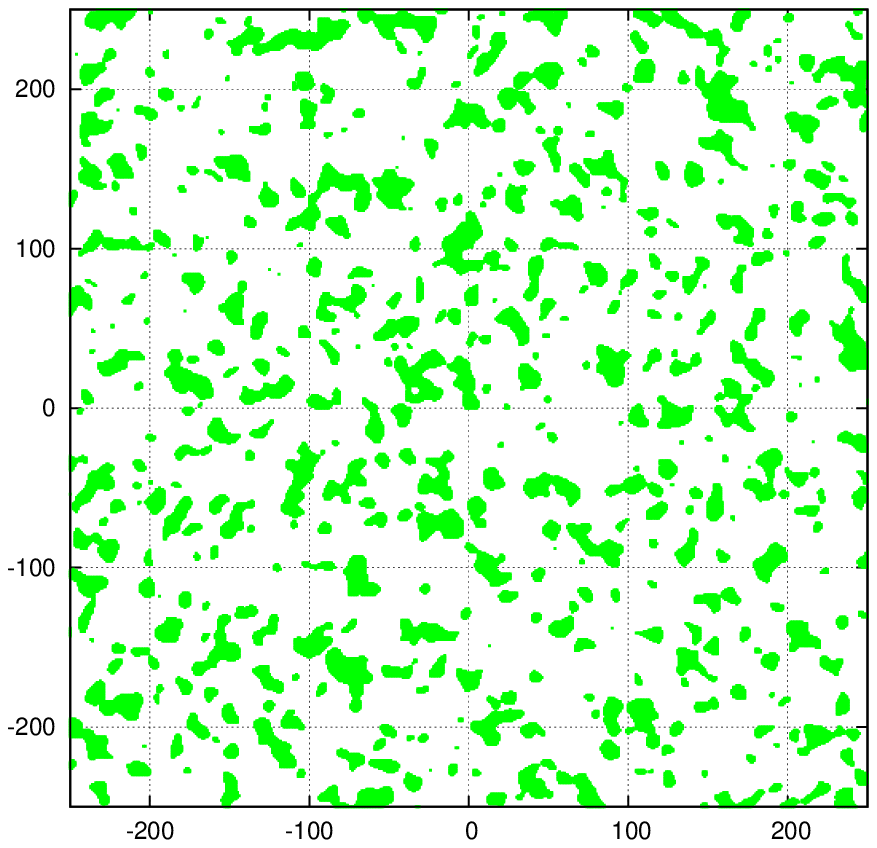,width=13cm} 
        \caption[Example of figure]{Lagrangian space. Regions $\delta(\vecq) > 1$ or equvalently $1 - I_1(\vecq) <0$ are shown in green.}%
	\label{fig:del-q}}

The specific volume is equal to the corresponding Jacobian
\be
{V_{\zeta}(\vecq,t) \over <{V}>} = J_{\zeta} = \det\left({\partial \zeta_i \over \partial q_j}\right).
\label{eq:vz}
\ee
The evaluation of $V_{\zeta}$ in terms of the initial fields is a quite laborious task.
Fortunately it allows a considerable simplification  in the limit of
$|D \vecs | \ll q$. Physically it means that the results are not reliable within 50 {\hmpc}
or so from the center but it is not very important since we are interested in much greater scales.
The components of the Jacobian become (to the zeroth order in
$| D \vecs |/q$)
\be
{\prtl \zeta_i \over \prtl q_j} \approx \delta_{ij} - D (d_{ij} + f  \tilde{n}_i \tilde{n}_k d_{jk}),
\ee
where $\tilde{\vecn}= \vecq/q$. 
Mapping to redshift space is neither irrotational nor solenodial
which means that  Jacobian $J_{\zeta}$ has no symmetry. Similarly to the Jacobian of the mapping
 to Eulerian space the determinant  $J_{\zeta}$ can be expressed  in terms of the invariants
 of the initial fields and  the radial components of two tensors $d_{qq}$  and $M_{qq}$
\ba
J_{\zeta} &=& 1 -D I_1 +D^2(1+f) I_2 - D^3(1+f) I_3
- fDd_{qq}  - f D^2 M_{qq},
\label{eq:Jz3}
\ea
where $d_{qq} = \tilde{n}_i \tilde{n}_j d_{ij}$ and $M_{qq} = \tilde{n}_i \tilde{n}_j M_{ij}$ 
are the radial components of the  deformation tensor and tensor made of the minors of
the determinant $I_3$. The  invariants are statistically isotropic fields although two of them ($I_2(\vecq)$ and $I_3(\vecq)$) are non-Gaussian. The last two terms depend on 
highly anisotropic fields: one $d_{qq}(\vecq)$ is Gaussian and  the other $M_{qq}(\vecq)$ is 
non-Gaussian. 
\section{Two-dimensional illustration}
\FIGURE{\epsfig{file=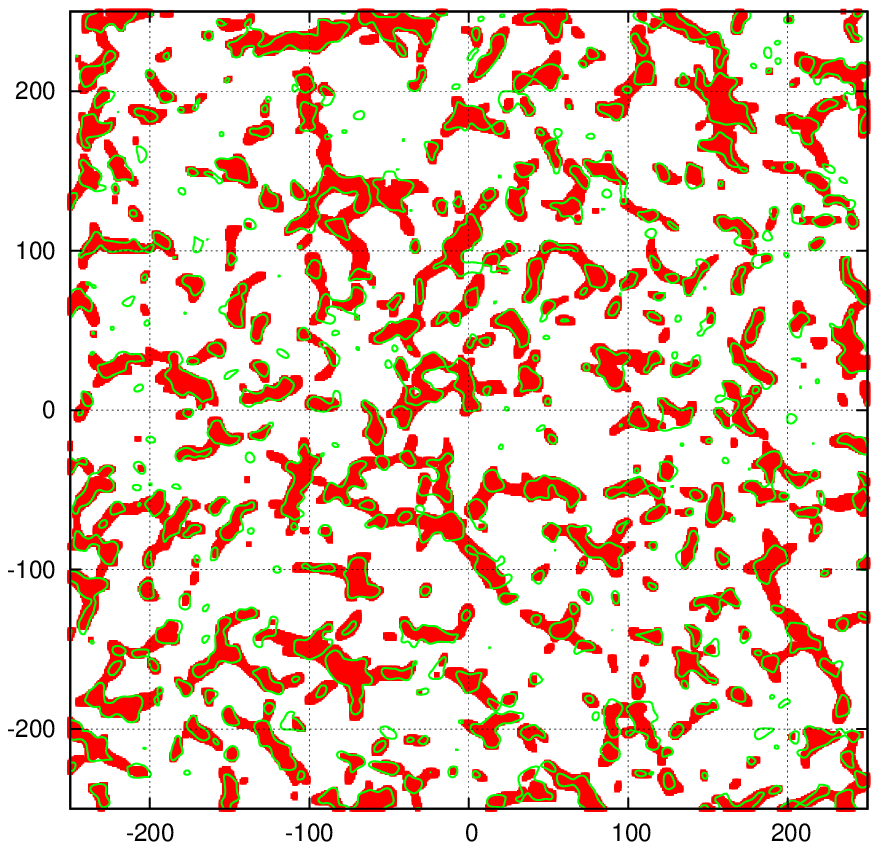,width=13cm} 
        \caption[Example of figure]{Lagrangian space. Red regions  $1- I_1(\vecq) + I_2(\vecq) < 0$, green contours   $\delta(\vecq)=1$ correspond to the boundaries of the regions in fig. \ref{fig:del-q}.}%
	\label{fig:jdel-q}}
Major although not all features of the model can be illustrated by a two-dimensional model.
Although the main goal of the paper is a study of the mapping to redshift space we
briefly discuss the mapping to real space and show the importance of non-Gaussian fields
for building the large-scale structure. We shall use a convenient normalization of function 
$D(t)$ such that at the present time $D(t_0)=1$ and  the variance of the linear density 
contrast field  $\sigma^2_{\delta} =1$, parameter $f \approx 0.8$ for the chosen cosmological
model.
\subsection{Mapping to real space}
The Jacobian of the mapping to Eulerian space 
 in two dimensions depends on two statistically isotropic  and 
homogeneous fields:  one Gaussian  $I_1(\vecq)$ and one non-Gaussian $I_2(\vecq)$ 
\be
{V_{x} \over <{V}>} = J_{x} = 1-DI_1+D^2I_2.
\label{eq:vx2}
\ee
The initial density contrast field $\delta_{in} \equiv \delta \rho/\bar{\rho} =
D(t_{in}) I_1$  is assumed to be Gaussian. 
Let us begin with the linear approximation ${V_{x} / <{V}>} = 1-DI_1$.
Linear theory formally predicts that the peaks with $\delta \ge 1$ 
shrink to the regions having formally negative volumes (and therefore negative density) by the present time.
Figure \ref{fig:del-q} shows an example of the progenitor of the structure 
in Lagrangian space predicted by the linear theory 
${V_{x}(t_0,q) / <{V}>} = 1- D(t_0) I_1(q) = 1 - I_1(q) < 0$. Although the initial power spectrum
is filtered with Gaussian windows $\exp(-k^2R_f^2/2)$ with $R_f =2.7$ {\hmpc} some peaks 
$\delta >1$ have quite large lengths reaching $\sim 30$ {\hmpc}.

The full nonlinear model 
$v_x(t_0,q) = 1- D(t_0) I_1(q)+ D^2(t_0) I_2(q) = 1- I_1(q)+ I_2(q) <0$
predicts the progenitor of the structure shown in Figure \ref{fig:jdel-q}  in red.
The green contours show the boundaries of the linear progenitor shown in
fig.\ref{fig:del-q}, thus one can see both the similarities and differences of two fields. 
In order to make an eye ball estimate of sizes easier we plot the grid on the background. 
The similarity of two fields
is obvious but it is not unexpected because both fields depend on  $I_1(q)$
and therefore strongly correlated.
The effect of non-Gaussian $I_2$-field is quite remarkable. The red 
non-Gaussian regions are considerably longer than the linear density peaks. 
They also typically comprise several green regions therefore non-Gaussian field $I_2$ provides
"bridges" linking the peaks  of linear density contrast field. 
This is in a good qualitative agreement with the 
observations made by Weinberg and Gunn  \cite{wei-gun90} who emphasized 
that the large structures usually consist of several peaks in the linear density field.
There is also no contradiction to the role of the correlation  between linear density peaks
in the BKP model \cite{bon-etal96}, although the the higher peaks
of the field smoothed with greater scale is assumed in the BKP model. 
\FIGURE{\epsfig{file=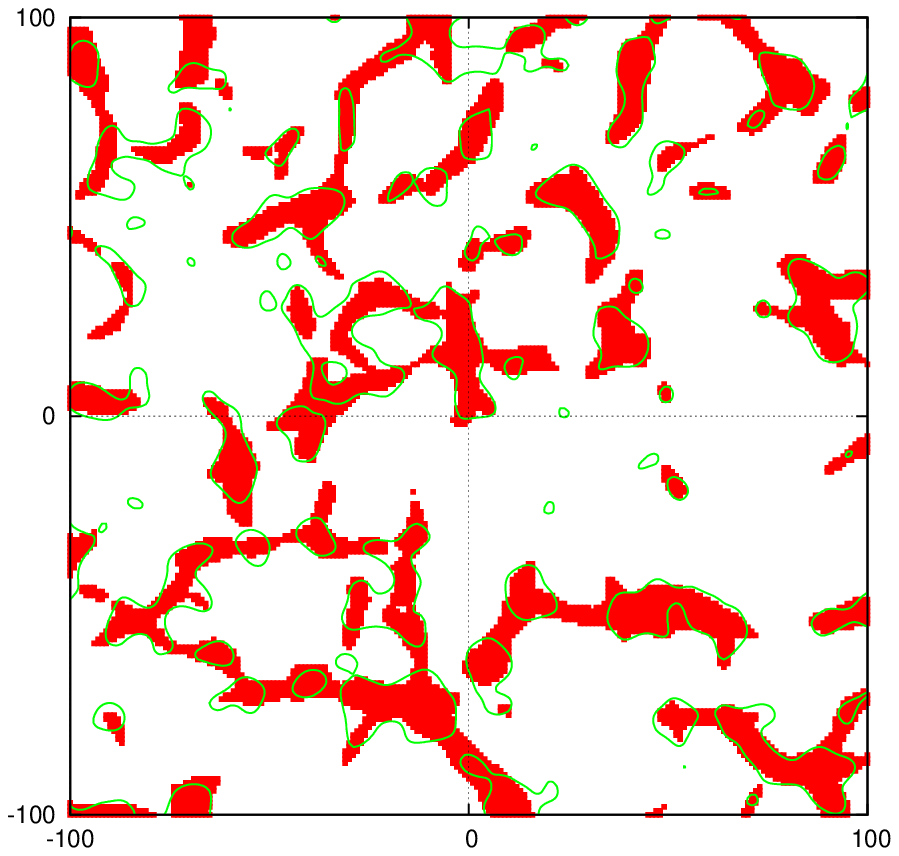,width=13cm} 
        \caption[Example of figure]{Lagrangian space, the central 200 {\hmpc} box of fig.\ref{fig:jdel-q}. Red regions  $1- I_1(\vecq) + I_2(\vecq) < 0$, green contours   $\delta(\vecq)=1$ correspond to the boundaries of the regions in fig.\ref{fig:del-q}.}%
	\label{fig:jdel-q-200}}
The  green contours totally comprise  15.9\%  of the area. It is remarkable
that  the red regions where the nonlinear condition  $1- I_1(q)+ I_2(q) <0$  
is satisfied occupy slightly less area (15.3\% ) than the green 
contours. Despite the greater lengths of the nonlinear progenitor
reaching over 100 {\hmpc} red regions are generally
slimmer than their counterparts in the linear progenitor. The resolution 
of fig.\ref{fig:jdel-q} is far too low to reveal this completely.  
However, it is obvious in fig.\ref{fig:jdel-q-200} where the central 
part of fig.\ref{fig:jdel-q} is zoomed in. In addition, small isolated regions of the linear
progenitor disappear from the nonlinear one (empty green contours).
\FIGURE{\epsfig{file=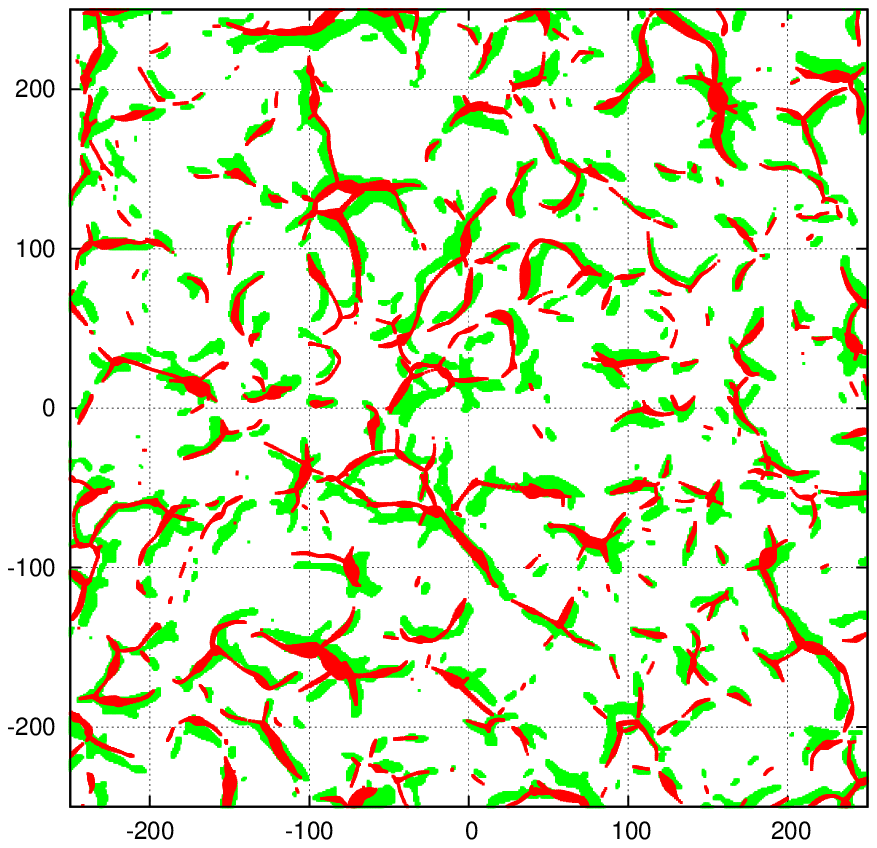,width=13cm} 
        \caption[Example of figure]{Eulerian space. The large-scale structure in Eulerian space (red) and the progenitor of the structure  in Lagrangian space (green).}%
	\label{fig:jdel-x}}
 
Figure \ref{fig:jdel-x} shows  the structure in Eulerian space (red) along with the
progenitor in Lagrangian space (green).  
The mapping makes the structure
considerably slimmer than its progenitor in  Lagrangian space although it is probably
thicker than in reality \cite{sh-z89}.  Two points are worth stressing. First, neither lengths
of the filaments no topology visibly change due to mapping itself. And second,
the filaments do not move much, however those that moved were are displaced 
mostly in the transverse direction while the longitudinal displacement is significantly smaller. 
It is also worth recalling that the {\rms} displacement of fluid elements
from the Lagrangian to Eulerian positions is about 15 {\hmpc} which greater than 
the thickness of the filaments but considerably smaller than their lengths.
The mapping to real space also involves statistically isotropic Gaussian vector 
field $s_i(\vecq) = - \prtl \Phi/ \prtl q_i$.
\subsection{Mapping to redshift space}
\smallskip
\FIGURE{\epsfig{file=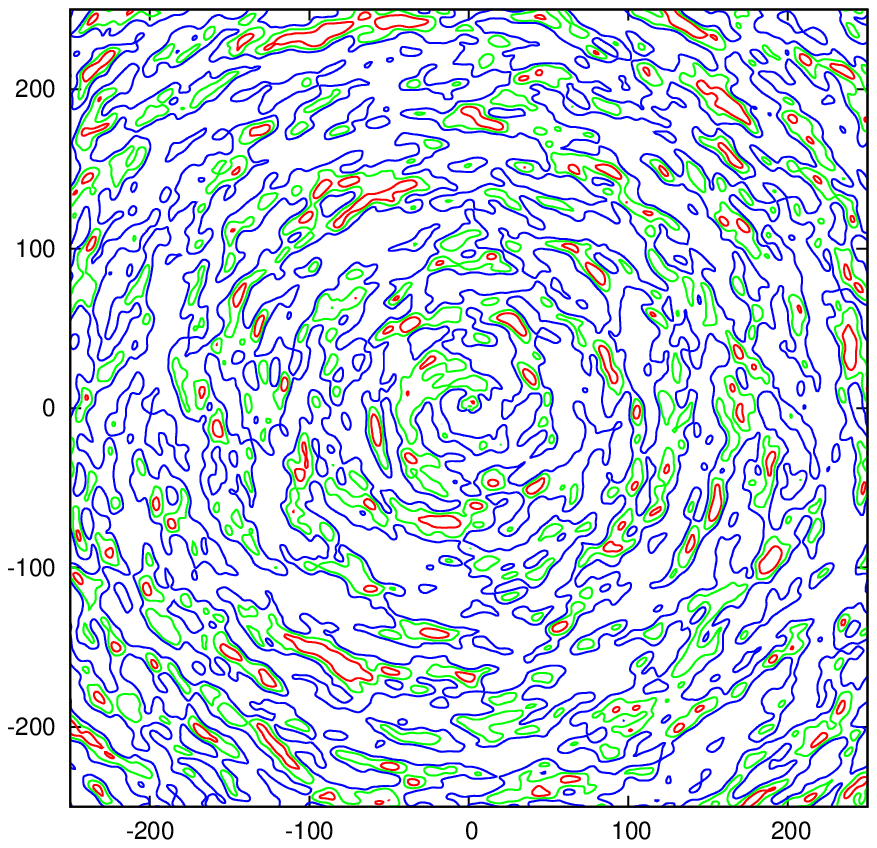,width=13cm} 
        \caption[Example of figure]{Lagrangian space. Contours  $f d_{qq}(\vecq) = 0.25, 0.5, 1$ are shown in blue, green and red respectively.}%
	\label{fig:dqq-q}}
\FIGURE{\epsfig{file=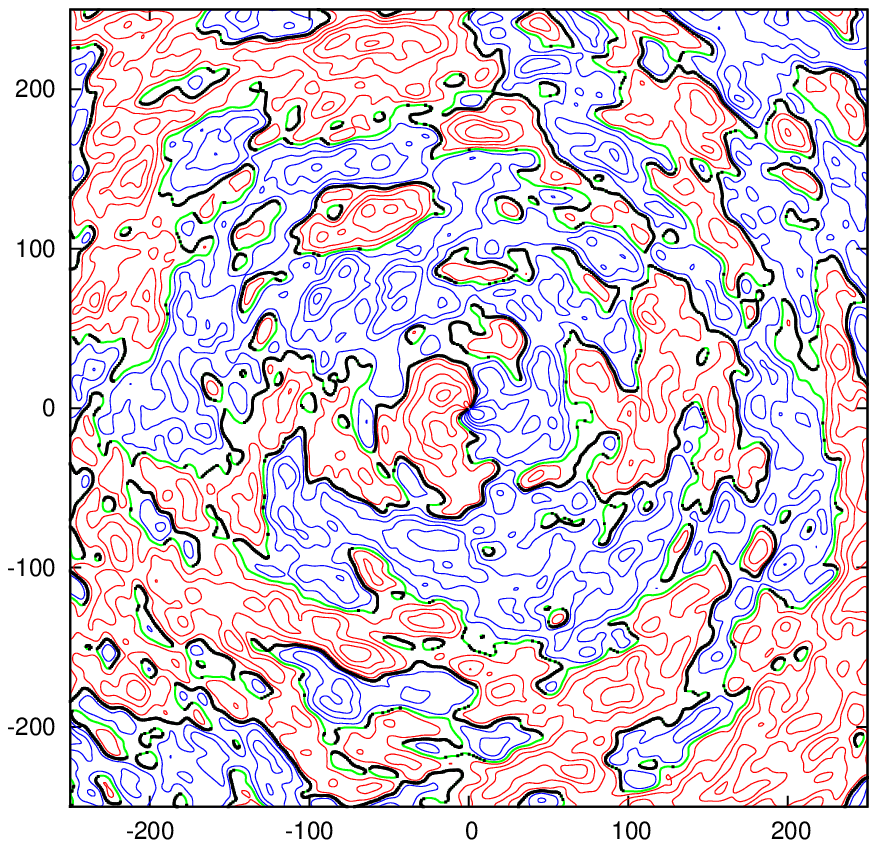,width=13cm} 
        \caption[Example of figure]{Lagrangian space. Contours of $f v_{q}(\vecq) >0$ are shown in red, $f v_{q}(\vecq) = 0$ in green and $f v_{q}(\vecq)<0 $ in blue. The segments of convergence {\ie} 
        the parts of the contour 
        $v_q(\vecq)=0$ satisfying the condition $\prtl v_q/ \prtl q <0$ are marked by black dots.}%
	\label{fig:vq-q}}
\FIGURE{\epsfig{file=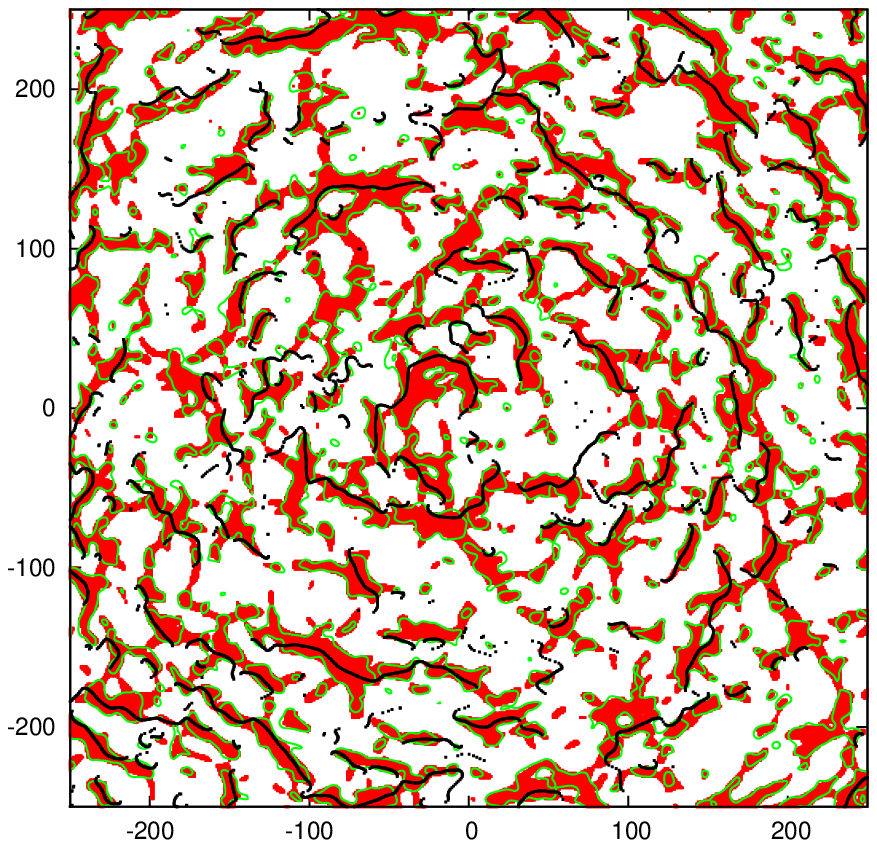,width=13cm} 
        \caption[Example of figure]{Lagrangian space.
The nonlinear progenitor of the structure in redshift space  defined by the condition eq. \ref{eq:Jz2} is shown in red. Green contours show the regions defined by the linear condition $\delta(\vecq) \ge 1$.
Black dots mark the segments of convergence.}%
	\label{fig:jz-q}}

We obtain the  Jacobian of the mapping to redshift space in two dimensions from
eq.  {\ref{eq:Jz3}
\be
J_{\zeta}(\vecq, t) = 1-DI_1+D^2(1+f) I_2 - fDd_{qq},
\label{eq:Jz2}
\ee
 where
 $I_1=d_{11}+d_{22}$ and $I_2=d_{11}d_{22}-d_{12}^2$ are the invariants of the deformation 
 tensor in two dimensions,
$d_{qq} = \tilde{n}_1^2 d_{11} + 2 \tilde{n}_1 \tilde{n}_2 d_{12} + \tilde{n}_2^2 d_{22}$
is the radial component of the deformation tensor.
Compared to the mapping to Eulerian space the term proportional to $I_2$ is boosted 
by factor $1+f$. The Jacobian contains a highly
anisotropic  field $d_{qq}(\vecq)$  in the sense that 
the contours are  significantly longer in the transverse  direction than in radial 
direction as a result the field acquires a characteristic circular pattern  (fig.\ref{fig:dqq-q}).  
The cause of this anisotropy becomes  much more obvious if one considers the small angle approximation \cite{kai87}.
Now, let us consider the direction corresponding  to $q_1$-axis. 
In this case  $\tilde{\vecn} = (1,0)$ therefore  
$d_{qq} =d_{11}(q) = \prtl^2 \Phi(\vecq)/\prtl q_1^2$. 
Differentiating potential two times with  respect to $q_1$ multiplies the power spectrum by $k_1^4$
($P_{d_{11}}(\veck) = k_1^4 P_{\Phi}(k)$) that effectively reduces the characteristic scale  of the resulting field along $q_1$ direction while
the scale along $q_2$ changes much less. In the case of arbitrary direction the reduction
of the scale along radial direction is much stronger than in the transverse direction that generates 
a distinct circular pattern.

The mapping to redshift space also involves the radial component of the velocity field 
$v_q(\vecq) = fDs_q(\vecq) =-fD \prtl \Phi/ \prtl q$ shown in fig.\ref{fig:vq-q}. 
Although the vector field $\vecs(\vecq)$ itself is isotropic its  components
are highly anisotropic fields. 
Again, let us consider the small angle approximation and $v_1(\vecq) \propto s_1(\vecq)$ 
component of the velocity vector field.   
The power spectrum
$P_{s_1}(\veck) = k_1^2 P_{\Phi}(k)$ is anisotropic with considerably more power along $k_1$ 
direction.
Therefore the field $s_1(\vecq)$ has much smaller scale along $q_1$ direction than 
along $q_2$ direction. For an arbitrary direction $\tilde{n}_i = q_i/q$ it is translated into 
considerable anisotropies between radial and transverse directions. 
A natural expectation suggests that the structure in redshift space is shifted to the segments 
of the zero contour $v_q(\vecq) =0$  satisfying the convergence condition 
$\prtl v_q / \prtl q <0$ marked by black dots on green contour  in fig.\ref{fig:vq-q}.
On the inner side of these segments the radial component of the peculiar velocity is positive and
therefore directed away from the center while on the outer side it is negative and thus directed
toward  the center resulting in shifting mass to these lines by the mapping to redshift space. 
\FIGURE{\epsfig{file=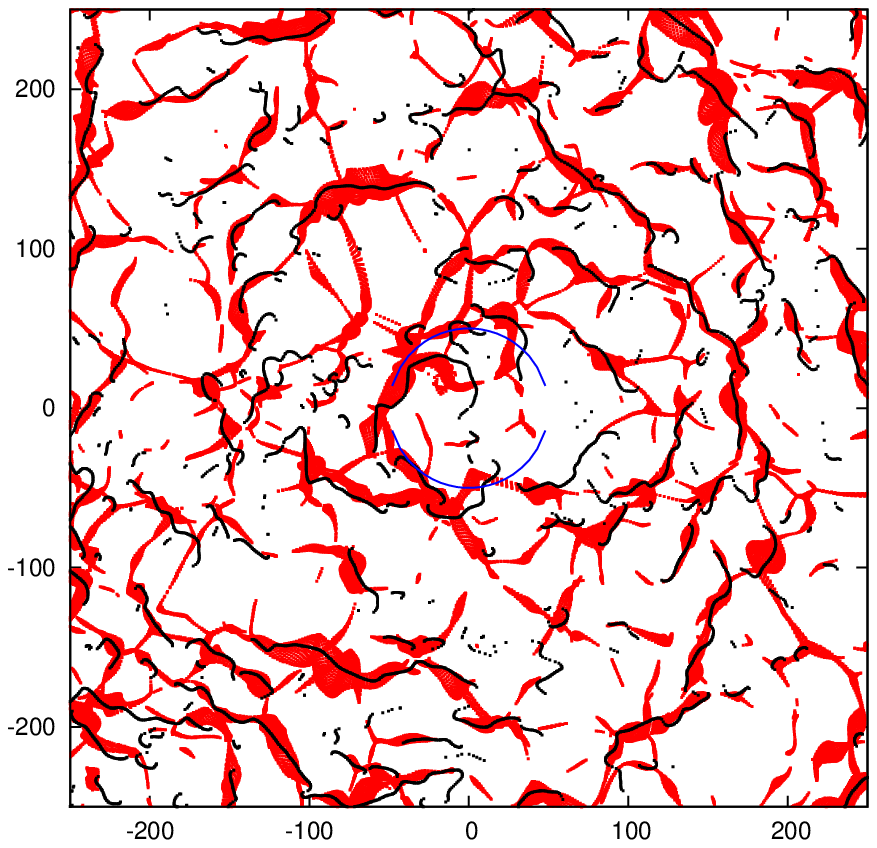,width=13cm} 
        \caption[Example of figure]{Redshift space. The large-scale structure is shown in red. The black lines are the parts of the contour $v_q(\vecq) \equiv v_i(\vecq)\tilde{n}_i(\vecq) =0$ selected by the condition $\prtl v_q /\prtl q <0$. Two blue arcs mark the central 50 {\hmpc} region where the adopted approximation 
($D |\vecs|/q \ll 1$ ) is not very accurate.}%
	\label{fig:jz-z}}

The progenitor of the structure in redshift space determined by eq. \ref{eq:Jz2}
is shown as red regions in  fig.\ref{fig:jz-q}. The green contours 
show the boundaries of the regions $\delta(\vecq) >1$ shown in fig.\ref{fig:del-q} as green 
regions and also as green contours in fig.\ref{fig:jdel-q}.
It is worth comparing the progenitors of the mappings to Eulerian (fig.\ref{fig:jdel-q}) 
and redshift space (fig.\ref{fig:jz-q}).  
The circular pattern due to $fDd_{qq}$ term in eq. \ref{eq:Jz2} is quite obvious in fig.\ref{fig:jz-q} 
but it is significantly less conspicuous than in fig.\ref{fig:dqq-q} because the other two fields
($I_1$ and $I_2$) affecting the Jacobian  are both statistically isotropic and homogeneous.
The black dots mark the lines of convergence defined above and shown in fig.\ref{fig:vq-q}.
The convergence lines obviously correlate with the progenitor of the mapping to redshift space
although the correlation is not perfect. 

The result of the mapping of the nonlinear progenitor (eq. \ref{eq:Jz2}) to  redshift space  
is shown  as red regions in fig.\ref{fig:jz-z}. 
If the linear progenitor was mapped to redshift space it would sit perfectly on top of the 
nonlinear  structure,  however  it  would remain significantly less connected than the nonlinear 
structure as  the mapping is continuous.
The figure illustrates a substantial change in the geometry caused by the mapping itself. 
The sizes of red
filaments and voids between them are noticeably greater in redshift space than 
in Lagrangian space.  
Some smaller regions seen in  Lagrangian space have merged into larger structures 
when mapped to redshift space. The difference between  the structures in Eulerian 
(fig.\ref{fig:jdel-x}) and redshift space (fig.\ref{fig:jz-z})  is quite substantial.

Figure \ref{fig:jz-z} also shows the convergence segments of the zero line of the radial 
velocity field  demonstrated in fig.\ref{fig:vq-q}.
Similarly to fig.\ref{fig:jz-q}  the black dots are noticeably more 
numerous in red regions than between them. 
In this case the correlation is not entirely perfect either but it is quite obvious despite 
the fact that the black dots are plotted in  Lagrangian
space while the red structure is plotted in redshift space. As we have seen in fig.\ref{fig:jdel-x}
the mapping to Eulerian space does not shift the progenitor much and the following
mapping to redshift space does not move the points with $v_q=0$.  In a few cases 
the separate segments of the convergence set seem to be aligned forming the filaments
with total length of over 500 {\hmpc}. An example of this is the system of five segments
that begins at distance about 200 {\hmpc} toward 1h from the center and continues 
clockwise down to distance  $\sim$150 {\hmpc} toward 4h from the center.

The correlation between the structure in redshift space and the segments of convergence
allows to use the lengths of the segment of convergence as  predictors of the length of
filaments.
Figure \ref{fig:conv-seg} shows the segments of convergence {\ie} the parts of line
$v_q(\vecq) = 0$ where additional conditions $ \prtl v_q/ \prtl q_q < 0 $ is satisfied.
These lines are attractors of the nonlinear structure in redshift space. Inspecting the 
figure one can easily  identify the lines extended over 500 {\hmpc} that suggests the
existence of the filaments of similar sizes. The field $s_q(\vecq)$ is closely related to
the initial potential since $s_q(\vecq) = -\prtl \Phi / \prtl q$ and therefore may be affected 
by an insufficient size of the box. Thus, the actual size of largest walls can be even greater than suggested by fig.\ref{fig:conv-seg} in real universe. The full three-dimensional N-body simulations
checking this prediction  are currently on the way.
\FIGURE{\epsfig{file=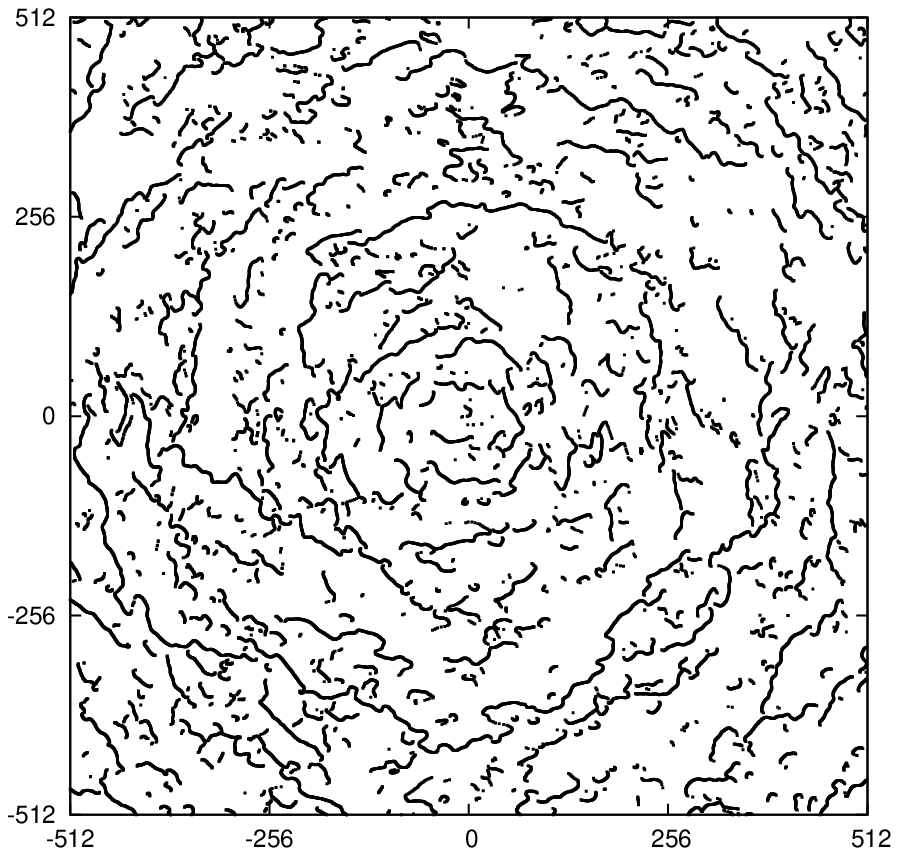,width=13cm} 
        \caption[Example of figure]{Lagrangian space (1024 {\hmpc} box).
The black dots mark the segments of convergence {\ie} the segments of contour $v_q(\vecq) = 0$
where additional condition $ \prtl v_q/ \prtl q_q < 0 $ is satisfied. These lines are a kind of attractor
of the structure in redshift space and their lengths are similar to the lengths of walls.}%
	\label{fig:conv-seg}}
\FIGURE{\epsfig{file=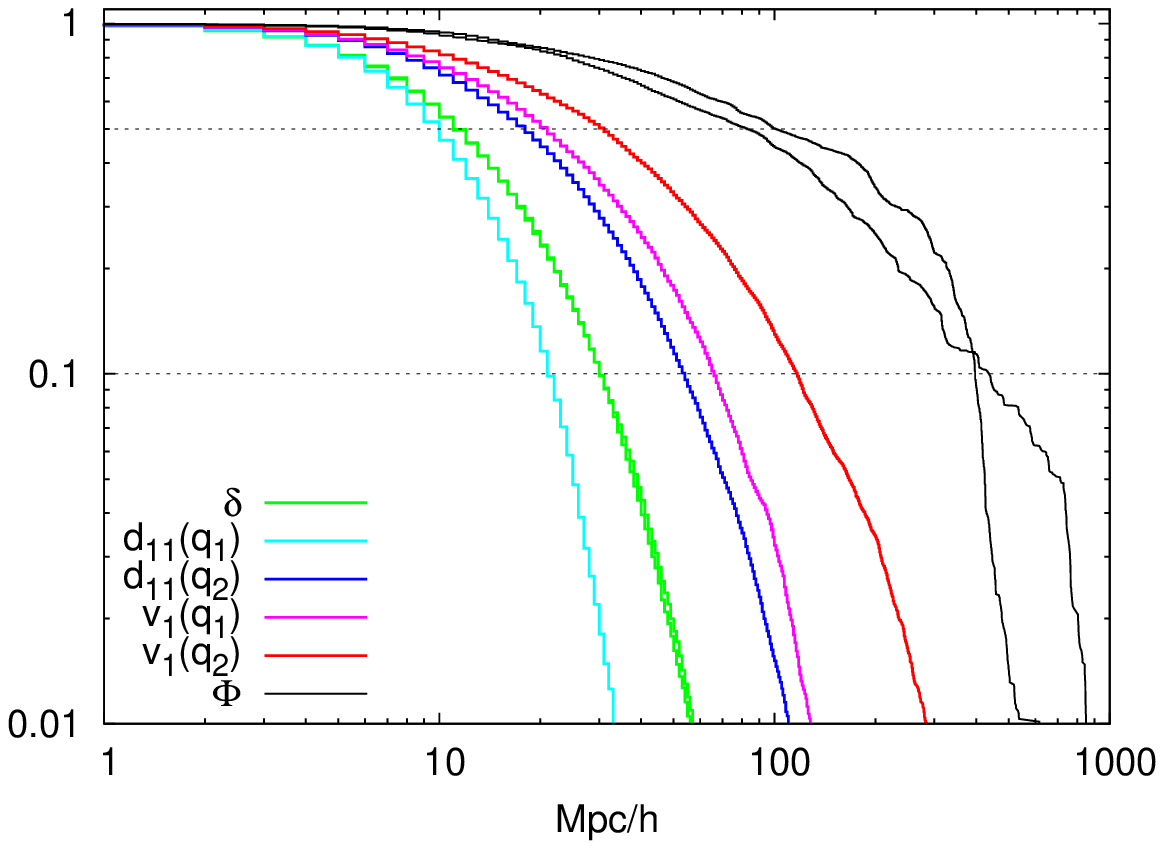,width=11cm} 
        \caption[Example of figure]{Cumulative probability function of the zero-crossing distances along two
axes $q_1$ and $q_2$. The top doted line marks the median values of the probability
functions. }%
	\label{fig:cross-stat}}
\subsection{Scales of Gaussian random fields}
In cosmology Gaussian  fields are often characterized by several scales derived from
moments $<k^{2j}> \equiv (2\pi^2)^{-1} \int k^{2j} P(k) k^2 dk k^{2j}$ of the power spectrum $P(k)$ ({\eg} \cite{bar-etal86}).
The ratios of these moments generate a set of scales
$ R_j^2 =  3 <k^{2j}> / <k^{2(j+1) }>$.
Two particular scales $R_1 \equiv R_*$ (using notation of ref. \cite{bar-etal86}) and  $R_0$ are used the most often
as the characteristics of  linear density fields. The former is associated with
the spatial distribution of peaks and is discussed in detail in \cite{bar-etal86}.
The latter is referred to as the mean distance between crossing points of
zero level \cite{ric54}. 
\cite{mon-yag75}  introduced the integral length scale
$L = \xi^{-1}(0) \int_{0}^{\infty} \xi(r) dr =(\pi/2) \int_{0}^{\infty} k P(k) dk / \int_{0}^{\infty} k^2 P(k) dk$, the physical meaning of which is not quite obvious. An example of using other scales in
cosmology is given in \cite{mou-tan05}. The suggested model
deals with the components of the deformation tensor $d_{ij}(\vecq)$ and displacement $s_i(\vecq)$
that are combined into radial components $d_{qq}(\vecq)$ and $s_q(\vecq)$. 
Here we consider the small angle approximation that fully described by $d_{11}$ and $s_1$
fields.
The scales can be characterized by the distribution of distances between successive zeros of a field
along two axes (see {\eg} \cite{mel-etal98}). Figure \ref{fig:cross-stat} shows the cumulative probability functions of zero crossing distances
for four fields ($\del = I_1, d_{11}$, $v_1$ and $\Phi$) along two axes ($q_1$ and $q_2$).
The density contrast  field is isotropic and both cpfs are practically identical. The potential field
also must be statistically isotropic but it depends strongly on long wave part of the power
spectrum. In fact the variance $\sigma^2_{\Phi}$ formally  logarithmically diverges at 
$k \rightarrow 0$. However, this is not a serious physical problem.  Heavens \cite{hea91} pointed
out that in a fixed sample volume (\eg the observable universe) the potential is well behaved,
however we cannot determine the global  zero level of the potential. He showed that by 
introducing $\Delta \Phi({\bf x}) = \Phi({\bf x}) - \Phi_0$, where $\Phi_0$ is the value of the potential 
at the observer location,  the problems related to the divergence of  $\sigma^2_{\Phi}$ are reduced
to a weak dependence of the results on unknown value of $\Phi_0$. Such fields are known as locally
homogeneous and locally isotopic fields and are discussed in Sec. 13 of ref. \cite{mon-yag75}.
This dependence of the potential on the long wave part of the spectrum explains 
 huge variations in the tail of the cumulative function seen in fig.\ref{fig:cross-stat} (black solid lines).
The scales along $q_2$ are much greater than along $q_1$ for both anisotropic 
fields  $d_{11}$ and $v_1$. The level lines of all fields are quite wiggly therefore there are
quite a few short distances between crossing points. The overall sizes are probably better
characterized by the tails of the distributions. Here are the median and the value marking
the top 10\% distances
along $q_1$ and $q_2$ axes for four fields in {\hmpc}: median
$\del$(11, 11),  $d_{11}$ (10, 18), $s_1$ (21, 31) and $\Phi$ (102, 84),
top 10\% $\del$ (30, 30),  $d_{11}$ (21, 54), $s_1$ (66, 117) and $\Phi$ (397, 437).
\section{Implications for three-dimensional universe}
The progenitor of the mapping to Eulerian space is determined by the interplay of  three fields:
one Gaussian $\delta = I_1$ and two non-Gaussian $I_2$ and $I_3$. It is easy to show
that they do not correlate by the direct calculation of the corresponding integrals
over the joint pdf (see Appendix). However the fact that two fields are non-Gaussian 
does not allow to declare them statistically independent and it makes the further 
analysis more challenging. 

Many features of two-dimensional model although not all are expected to be very similar 
in three dimensions.
Nonlinear term proportional to $I_2$   helps to fill the gaps between peaks of 
linear field $\delta = I_1(\vecq) > 1$ in 2D and therefore boosts the length of  filaments 
in Lagrangian space up to 70 --100 {\hmpc}. There is no reason to doubt
that it works in a similar way in 3D. In addition, the term proportional $I_3$ in
eq. \ref{eq:Jx3} is likely to boost the lengths of filaments even more.
However, the further study is obviously needed both analytical and numerical.

The mapping to Eulerian space itself does not change significantly the lengths of
the filaments and does not displace them much from the initial positions in 2D.
The mapping in 3D is expected to be similar in this respect.

However, the mapping to redshift space  differs in 3D significantly in a few respects.
All three fields $d_{qq}$, $M_{qq}$  (see eq. \ref{eq:Jz3}) and $s_q$ (see eq. \ref{eq:qtoz3}) 
have scales  in the radial direction much shorter
than in two transverse directions. This causes the progenitor and the structure in redshift 
space to acquire a circular pattern in 2D while it gives rise to the origin of shell like walls in 3D. 
The walls in redshift space expected to
be neither homogeneous no thin but their transverse sizes must be considerably greater than
the radial thickness. 
Similarly to the  two-dimensional case the greatness of 'great walls' comes from anisotropy of two
Gaussian fields $d_{qq}$ and $s_q$ and in addition one non-Gaussian field, $M_{qq}$
derived from initial Gaussian field. As fig. \ref{fig:cross-stat} indicates the transverse coherence
of both $d_{11}$ and  especially $s_1$ field can easily exceed 100 {\hmpc}.
Two isotropic non-Gaussian fields $I_2$ and $I_3$ also 
contribute to the  formation of the structure in redshift space since the corresponding terms are
boosted by factor $1+f$ in the Jacobian $J_{\zeta}$ (eq. \ref{eq:Jz3}). 

The convergence lines defined in 2D become the convergence surfaces in 3D, they are expected
to provide a reasonable estimate of the transverse sizes of great walls.
\section{Summary and discussion}
We propose a new theoretical model that describes the formation of the
large-scale structure in redshift space.
It naturally explains the sizes of the observed great walls and predicts that even greater
walls will be discovered in future deeper and larger redshift surveys. Although
based on the Zel'dovich approximation \cite{z70} the suggested model uses it in a
new and very different way. Instead of traditional approaches that focused on the evaluation 
of density or positions and velocities of the particles we introduce the notion of the
progenitor of the large-scale structure.  It is defined as a set of regions in Lagrangian space
that after mapping to Eulerian or redshift space (depending on the kind of mapping we study)  
end up in the regions where the particle trajectories have crossed and the mapping 
is no longer unique. Neglecting the thermal velocities and assuming smooth initial condition
the boundaries of these regions in Eulerian or redshift space are caustics. In reality due to
hierarchical clustering process the caustics do not form but the density of both matter and galaxies
is still highest in these regions ({eg} \cite{lit-wei-par91}). 
Even in the case of smooth initial conditions the evaluation of the density
within the shell-crossing regions is not easy and we do not attempt to estimate it in this paper. 
Instead we utilize the Zel'dovich approximation for identification of
the progenitor of the structure in Lagrangian space and then for  mapping it to Eulerian or 
redshift space. In particular we  demonstrate the role of various fields some of which are
non-Gaussian or/and anisotropic. This is the most important factor in explanation of the
lengths of filaments in Eulerian space. Non-Gaussian $I_2(\vecq)$ and $I_3(\vecq)$ 
fields fill the gaps between peeks of Gaussian $\delta(\vecq) = I_1(\vecq)$ field significantly 
boosting the length of the progenitor of the filaments in Lagrangian space. 
It is worth stressing that 
the mapping of the progenitor to Eulerian space does not increase its lengths noticeably
although it significantly affects its appearance.

The mapping to redshift space has  both the progenitor in Lagrangian 
space and the structure in redshift space significantly different than the mapping to 
Eulerian space. 
In both cases the thickness of the filaments in Eulerian space and walls
in redshift space is probably the least accurate in this approximation. The thickness
of filaments in Eulerian space is probably exaggerated \cite{sh-z89}. The thickness
of the walls in redshift space may be affected by nonlinear effects that change the velocities
in the shell-crossing regions. However, the estimate of the extent of the walls in 
the transverse direction is expected to be quite robust.

Although both the suggested model and one suggested by \cite{bon-etal96} (BKP)
are based on the Zel'dovich approximation (eq. \ref{eq:zappr}) there are a few radical
differences  between them.  The current model can explain the sizes of both the structures
in Eulerian and redshift space emphasizing that the structure in redshift space must have 
significantly greater sizes in transverse directions than in the line of sight direction and
therefore the wall kind structures must be more conspicuous in redshift space. 
The BKP model discusses only the structure in Eulerian space.
The BKP model uses the smoothing scale $R_b$ defined by an inequality
$\sigma_{\delta}(R_b ) \lesssim 1$ such that it corresponds to "the conceptual picture of a smooth
stately large-scale single-stream flow" \cite{bon-etal96} while the current model uses the scale of nonlinearity and therefore assumes the crossing of trajectories and presence of multi-stream flows.
Although it is not excluded that the further comparison with cosmological N-body simulations 
may slightly change the scale for better fit to N-body simulations it will not change the basic 
construction used by the model - the progenitor of the structure.  The model defines 
the progenitor of the structure  as a set of regions in Lagrangian space where 
the Jacobian of the mapping becomes negative indicating that these regions 
end up in multi-stream flow regions after mapping to Eulerian or redshift space. 
There is also one more principal difference. 
The BKP model uses only linear {\ie} Gaussian fields and conditional
correlations between them. 
This model utilizes both  non-Gaussian fields ($I_2, I_3, M_{qq}$) one of which 
is anisotropic as well as anisotropic Gaussian fields $d_{qq}, s_q$ along with linear density 
contrast field. The displacement fields $s_i$ and $s_q$  also play important roles in the 
appearances  of the structure in Eulerian and redshift spaces. 
The suggested approach allows easily to isolate the
role of every field and thus helps to better understand the complex nonlinear dynamics
of the structure formation.

Further studies including the more detailed studies of the statistical properties of the fields
involved in the mappings as well as a direct comparison with cosmological N-body simulations
are obviously needed. Some work has already started and the results will be reported
in the following papers.

\acknowledgments
I would like to acknowledge many useful discussions I had during last several years
with J. R. Bond,  U. Frisch, S. Habib, K. Heitmann, L. Kofman, A. S. Szalay, R. Triay, 
B. Tully, and R. van de Weygaert that helped me greatly  in developing this model.
I am also grateful to the anonymous referee for constructive and useful comments.

\appendix
\section{ Appendix: PDF of invariants $I_1, I_2$ and $I_3$}
From eq.  12 in \cite{lee-sh98} it is easy to derive the joint distribution of three invariants
if one notes that 
$dI_1\,dI_2\, dI_3 = (\lambda_1 -\lambda_2)(\lambda_1 -\lambda_3)(\lambda_2 -\lambda_3) d\lambda_1\, d\lambda_2\,d\lambda_3$ ($\lambda_i$ and $I_1$ are measured in units of
$\sigma_{\delta}$, $I_2$  in units of $\sigma_{\delta}^2$ and $I_3$  in units of $\sigma_{\delta}^3$)
\be
p(I_1,I_2,I_3) = {675\sqrt{5}  \over 8\pi}\exp\left(- {3I_1^2} \right) F_2(I_1,I_2)\, F_3(I_1,I_2,I_3),
\ee
where
\ba
F_2(I_1,I_2) &=&\exp\left({15 \over 2} I_2\right),~~ {\rm if}~~I_2<I_1^2/3
~~{\rm and}~~0~~{\rm otherwise}, \nonumber  \\ 
F_3(I_1,I_2, I_3)&=& 1,~~~~~~~~~~~~~~~~~{\rm if}~~I_3^{(2)} < I_3 < I_3^{(1)} 
~~{\rm and}~~0~~{\rm otherwise}  \nonumber
\ea
The constraints on $I_3$ reflect the fact that all three $\lambda_i$ 
are real which requires that $\Delta(I_1,I_2,I_3) < 0$, where
\be
\Delta(I_1,I_2,I_3)= 27 I_3^2 - 2I_1(9I_2 - 2I_1^2)I_3 + I_2^2(4 I_2-I_1^2).
\ee
Therefore,  $I_3$ must lie between the roots of the equation $\Delta(I_1,I_2,I_3)=0$ 
which are
\be
I_3^{(1,2)} = {1 \over 27}\left[I_1(9I_2-2I_1^2)\pm 2(I_1^2-3I_2)^{3/2}\right].
\ee
The constraint on $I_2$ is a trivial consequence of relations between
the eigen values $\lambda_1, \lambda_2$ and $\lambda_3$ and
 the invariants: $I_1 = \lambda_1 +\lambda_2 +\lambda_3$, 
 $I_2=\lambda_1 \lambda_2+\lambda_1 \lambda_3 + \lambda_2 \lambda_3 $
 and $I_3=\lambda_1 \lambda_2 \lambda_3$.
Thus, the distribution of $I_3$ for given $I_1$ and $I_2$  is uniform in the range $I_3^{(2)} < I_3 < I_3^{(1)}$.  The distribution of $I_2$ for given $I_1$ and $I_3$
is exponential in the range $-\infty < I_2<I_1^2/3$ and the distribution of $I_1$
for given $I_2$ and $I_3$ is Gaussian.
Integration over $I_3$ gives the joint distribution of $I_1$ and $I_2$
\be
p(I_1,I_2) = {25 \sqrt{5} \over 2 \pi} \exp\left(-{3I_1^2}\right)
\exp\left({15 \over 2} I_2 \right) (I_1^2-3I_2)^{3/2},
\ee
and then the integration over $I_2$ gives the Gaussin distribution of the
linear density contrast since $\delta = D I_1$ in the linear regime.
It is not difficult to check that the invariants do not correlate at one point: \ie
$<I_1I_2> =0$, $<I_1I_3> =0$ and $<I_2I_3> =0$.

\end{document}